\documentclass[12pt,a4paper]{article}

\usepackage{lineno}  % for line numbering during review
\usepackage{graphicx}  % to include figures (can also use other packages)
\usepackage{xspace} % To avoid problems with missing or double spaces after
\usepackage{color}
\usepackage{colortbl}
\usepackage{amsmath} % Adds a large collection of math symbols
\usepackage{ifthen} % for conditional statements

\usepackage{amsfonts}

\usepackage{booktabs}
\usepackage{multirow}
\usepackage{times}
\usepackage{mathptmx}
\usepackage{rotating}
\usepackage[T1]{fontenc}

\newboolean{pdflatex}
\setboolean{pdflatex}{true} % use this if using non-eps figures
\newboolean{articletitles}
\setboolean{articletitles}{true} % False removes titles in references

\newboolean{uprightparticles}
\setboolean{uprightparticles}{false} %Set to true to get roman particle symbols

\usepackage{amssymb}
\usepackage{amsfonts}
\usepackage{upgreek} % Adds in support for greek letters in roman typeset
\usepackage{mciteplus}
\usepackage{hyperref}    % Hyperlinks in references
\usepackage[all]{hypcap} % Internal hyperlinks to floats.

\DeclareMathAlphabet{\mathbfi}{OT1}{ptm}{bx}{it}
\def\lhcb {LHCb\xspace}
\def\ux85 {UX85\xspace}

\ifthenelse{\boolean{uprightparticles}}%
{

 \def\Ppsi        {\ensuremath{\uppsi}\xspace}

 \def\PDelta      {\ensuremath{\Delta}\xspace}                 
 \def\PXi      {\ensuremath{\Xi}\xspace}                 
 \def\PLambda      {\ensuremath{\Lambda}\xspace}                 
 \def\PSigma      {\ensuremath{\Sigma}\xspace}                 
 \def\POmega      {\ensuremath{\Omega}\xspace}                 
 \def\PUpsilon      {\ensuremath{\Upsilon}\xspace}                 
 
 \def\PB      {\ensuremath{\mathrm{B}}\xspace}                 
                  
 \def\PD      {\ensuremath{\mathrm{D}}\xspace}

 \def\PJ      {\ensuremath{\mathrm{J}}\xspace}                 
 \def\PK      {\ensuremath{\mathrm{K}}\xspace}

 \def\Pb      {\ensuremath{\mathrm{b}}\xspace}                 
 \def\Pc      {\ensuremath{\mathrm{c}}\xspace}

 \def\Pi      {\ensuremath{\mathrm{i}}\xspace}

}
{

 \def\Ppsi        {\ensuremath{\psi}\xspace}                 
                  
 \mathchardef\PDelta="7101
 \mathchardef\PXi="7104
 \mathchardef\PLambda="7103
 \mathchardef\PSigma="7106
 \mathchardef\POmega="710A
 \mathchardef\PUpsilon="7107
                  
 \def\PB      {\ensuremath{B}\xspace}                 
                  
 \def\PD      {\ensuremath{D}\xspace}

 \def\PJ      {\ensuremath{J}\xspace}                 
 \def\PK      {\ensuremath{K}\xspace}

 \def\Pb      {\ensuremath{b}\xspace}                 
 \def\Pc      {\ensuremath{c}\xspace}

 \def\Pi      {\ensuremath{i}\xspace}

}

\def\cquark    {\ensuremath{\Pc}\xspace}

\def\bquark    {\ensuremath{\Pb}\xspace}
\def\bquarkbar {\ensuremath{\overline \bquark}\xspace}
\def\bbbar     {\ensuremath{\bquark\bquarkbar}\xspace}

\def\kaon  {\ensuremath{\PK}\xspace}
  \def\Kbar  {\kern 0.2em\overline{\kern -0.2em \PK}{}\xspace}

\def\Kz    {\ensuremath{\kaon^0}\xspace}
\def\Kzb   {\ensuremath{\Kbar^0}\xspace}
\def\KzKzb {\ensuremath{\Kz \kern -0.16em \Kzb}\xspace}
\def\Kp    {\ensuremath{\kaon^+}\xspace}
\def\Km    {\ensuremath{\kaon^-}\xspace}

\def\KpKm  {\ensuremath{\Kp \kern -0.16em \Km}\xspace}

  \def\Dbar    {\kern 0.2em\overline{\kern -0.2em \PD}{}\xspace}
\def\D       {\ensuremath{\PD}\xspace}

\def\Dz      {\ensuremath{\D^0}\xspace}
\def\Dzb     {\ensuremath{\Dbar^0}\xspace}
\def\DzDzb   {\ensuremath{\Dz {\kern -0.16em \Dzb}}\xspace}
\def\Dp      {\ensuremath{\D^+}\xspace}
\def\Dm      {\ensuremath{\D^-}\xspace}

\def\DpDm    {\ensuremath{\Dp {\kern -0.16em \Dm}}\xspace}

\def\B       {\ensuremath{\PB}\xspace}
  \def\Bbar    {\kern 0.18em\overline{\kern -0.18em \PB}{}\xspace}

\def\Bpm     {\ensuremath{\B^\pm}\xspace}

\def\jpsi     {\ensuremath{{\PJ\mskip -3mu/\mskip -2mu\Ppsi\mskip 2mu}}\xspace}

  \def\Y#1S{\ensuremath{\PUpsilon{(#1S)}}\xspace}% no space before {...}!

\def\to                 {\ensuremath{\rightarrow}\xspace}

\def\AT#1     {\ensuremath{A_{\mathrm{T}}^{#1}}\xspace}           % 2

\def\C#1      {\ensuremath{\mathcal{C}_{#1}}\xspace}                       % 9
\def\Cp#1     {\ensuremath{\mathcal{C}_{#1}^{'}}\xspace}                    % 7
\def\Ceff#1   {\ensuremath{\mathcal{C}_{#1}^{\mathrm{(eff)}}}\xspace}        % 9  
\def\Cpeff#1  {\ensuremath{\mathcal{C}_{#1}^{'\mathrm{(eff)}}}\xspace}       % 7
\def\Ope#1    {\ensuremath{\mathcal{O}_{#1}}\xspace}                       % 2
\def\Opep#1   {\ensuremath{\mathcal{O}_{#1}^{'}}\xspace}                    % 7

\newcommand{\tev}{\ensuremath{\mathrm{\,Te\kern -0.1em V}}\xspace}
\newcommand{\gev}{\ensuremath{\mathrm{\,Ge\kern -0.1em V}}\xspace}
\newcommand{\mev}{\ensuremath{\mathrm{\,Me\kern -0.1em V}}\xspace}
\newcommand{\kev}{\ensuremath{\mathrm{\,ke\kern -0.1em V}}\xspace}
\newcommand{\ev}{\ensuremath{\mathrm{\,e\kern -0.1em V}}\xspace}
\newcommand{\gevc}{\ensuremath{{\mathrm{\,Ge\kern -0.1em V\!/}c}}\xspace}
\newcommand{\mevc}{\ensuremath{{\mathrm{\,Me\kern -0.1em V\!/}c}}\xspace}
\newcommand{\gevcc}{\ensuremath{{\mathrm{\,Ge\kern -0.1em V\!/}c^2}}\xspace}
\newcommand{\gevgevcccc}{\ensuremath{{\mathrm{\,Ge\kern -0.1em V^2\!/}c^4}}\xspace}
\newcommand{\mevcc}{\ensuremath{{\mathrm{\,Me\kern -0.1em V\!/}c^2}}\xspace}

\def\mum  {\ensuremath{\,\upmu\rm m}\xspace}

\def\mub{\ensuremath{\rm \,\upmu b}\xspace}

\def\invpb {\ensuremath{\mbox{\,pb}^{-1}}\xspace}

\def\gsim{{~\raise.15em\hbox{$>$}\kern-.85em
          \lower.35em\hbox{$\sim$}~}\xspace}
\def\lsim{{~\raise.15em\hbox{$<$}\kern-.85em
          \lower.35em\hbox{$\sim$}~}\xspace}

\def\tell1  {TELL1\xspace}
\def\ukl1   {UKL1\xspace}

\newcommand{\pbinv}{\invpb}

\newcommand{\ptrans}{\ensuremath{p_{\rm T}}}

\newcommand{\bplus}{\Bpm}

\newcommand{\bpjpsik}{\ensuremath{\bplus\to\jpsi K^\pm}}
\newcommand{\bpjpsipi}{\ensuremath{\bplus\to\jpsi \pi^\pm}}
\newcommand{\jpsimumu}{\ensuremath{\jpsi\to\mu^+ \mu^-}}

\newcommand{\xsresult}{\ensuremath{41.4\pm 1.5\,({\rm stat.}) \pm 3.1\,({\rm syst.}) \,\mub}}
\graphicspath{{figs/}}

\usepackage{mciteplus}

\begin{document}
%\setpagewiselinenumbers
%\modulolinenumbers[2]
%\switchlinenumbers
%\linenumbers
%%%%%%%%%%%%%%%%%%%%%%%%%
%%%%% Title     %%%%%%%%%
%%%%%%%%%%%%%%%%%%%%%%%%%
\renewcommand{\thefootnote}{\fnsymbol{footnote}}
\setcounter{footnote}{1}

% %%%%%%% CHOOSE --------
% $Id: title-LHCb-PAPER.tex 16067 2012-02-21 00:08:05Z jhe $
% ===============================================================================
% Purpose: LHCb-PAPER journal paper title page template
% Author:
% Created on: 2010-09-25
% ===============================================================================

%%%%%%%%%%%%%%%%%%%%%%%%%
%%%%%  TITLE PAGE  %%%%%%
%%%%%%%%%%%%%%%%%%%%%%%%%
\begin{titlepage}
\pagenumbering{roman}

% Header ---------------------------------------------------
\vspace*{-1.5cm}
\centerline{\large EUROPEAN ORGANIZATION FOR NUCLEAR RESEARCH (CERN)}
\vspace*{1.5cm}
\hspace*{-0.5cm}
\begin{tabular*}{\linewidth}{lc@{\extracolsep{\fill}}r}
\ifthenelse{\boolean{pdflatex}}% Logo format choice
{\vspace*{-2.7cm}\mbox{\!\!\!\includegraphics[width=.14\textwidth]{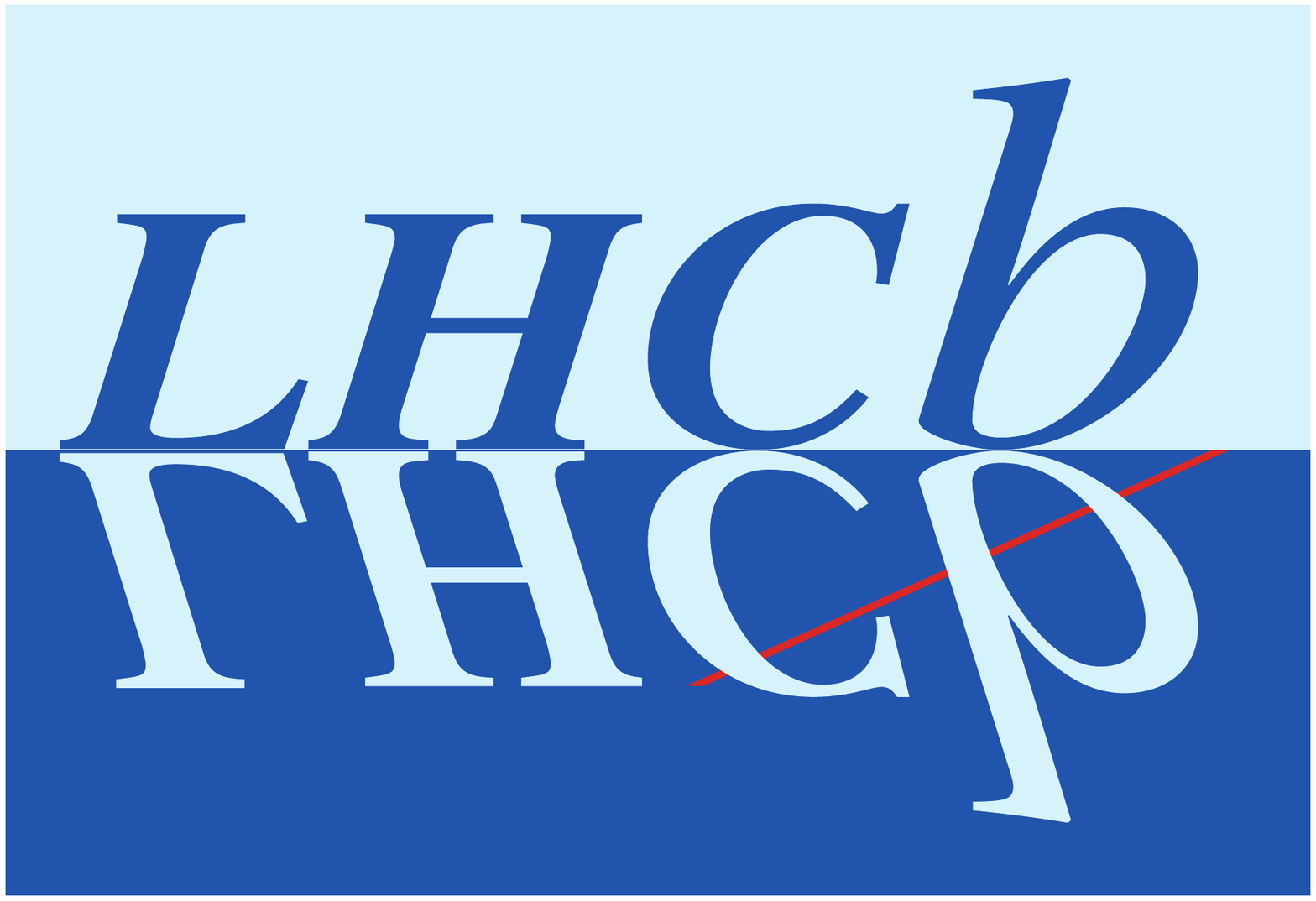}} & &}%
{\vspace*{-1.2cm}\mbox{\!\!\!\includegraphics[width=.12\textwidth]{figs/lhcb-logo.eps}} & &}%
\\
 & & CERN-PH-EP-2012-047 \\  % ID
 & & LHCb-PAPER-2011-043 \\  % ID
 & & \today \\ % Date - Can also hardwire e.g.: 23 March 2010
 & & \\
% not in paper \hline
\end{tabular*}

\vspace*{4.0cm}

%\vspace*{2.0cm}
% Title --------------------------------------------------
{\bf\boldmath\huge
\begin{center}
Measurement of the $\mathbfi{B^\pm}$ production cross-section in $\mathbfi{pp}$ collisions at $\mathbfi{\sqrt{s}=7}$ TeV
\end{center}
}

\vspace*{2.0cm}

% Authors -------------------------------------------------
\begin{center}
The LHCb collaboration
\footnote{Authors are listed on the following pages.}
\end{center}

\vspace{\fill}

% Abstract -----------------------------------------------
\begin{abstract}
  \noindent
  The production of $\bplus$ mesons in proton-proton collisions
  at $\sqrt{s}=7\,\tev$ is studied using 35 pb$^{-1}$ of
  data collected by the LHCb detector.
  The \bplus\ mesons are reconstructed exclusively in the \bpjpsik\ mode,
  with \jpsimumu.
  The differential production cross-section is measured as
  a function of the $\bplus$ transverse momentum
  in the fiducial region \mbox{$0<p_{\rm T}<40$ GeV/$c$}
  and with rapidity \mbox{$2.0<y<4.5$}.
  The total cross-section, summing up $B^+$ and $B^-$,
  is measured to be
  \begin{center}
    $\sigma(pp \to \bplus X,\;\mbox{$0<p_{\rm T}<40$\; GeV/$c$},\; 2.0<y<4.5) = \xsresult$.
  \end{center}
\end{abstract}

\begin{center}
Submitted to JHEP
\end{center}
\vspace*{2.0cm}
\vspace{\fill}

\end{titlepage}

%%%%%%%%%%%%%%%%%%%%%%%%%%%%%%%%
%%%%%  EOD OF TITLE PAGE  %%%%%%
%%%%%%%%%%%%%%%%%%%%%%%%%%%%%%%%

%  empty page follows the title page ----
\newpage
\setcounter{page}{2}
\mbox{~}
\newpage

% Author List ----------------------------
%  You need to get a new author list!
% \documentclass[a4paper]{article}
% \setlength{\oddsidemargin}{0cm}
% \setlength{\evensidemargin}{0cm}
% \setlength{\textwidth}{16.5cm}
% \setlength{\parindent}{0cm}
% \begin{document}
\centerline{\large\bf The LHCb collaboration} 
\begin{flushleft}
% {\Large LHCb Collaboration ----- official authorship list}\\[4ex]
% valid for date: 18. Dec. 2011\\
% used for paper: $B^+$ production (LHCb-PAPER-2011-043)\\[4ex]
% collaborators included, who did not leave before 18. Dec. 2010\\
%                            and who joined before 18. Jun. 2011\\[2ex]
% {\small today is 5. Feb. 2012}\\[4ex]
%-- 
%-- LHCb Authorlist, Status of 18. Dec. 2011
%-- 
R.~Aaij$^{38}$, 
C.~Abellan~Beteta$^{33,n}$, 
B.~Adeva$^{34}$, 
M.~Adinolfi$^{43}$, 
C.~Adrover$^{6}$, 
A.~Affolder$^{49}$, 
Z.~Ajaltouni$^{5}$, 
J.~Albrecht$^{35}$, 
F.~Alessio$^{35}$, 
M.~Alexander$^{48}$
% ...

{\footnotesize \it
$ ^{1}$Centro Brasileiro de Pesquisas F\'{i}sicas (CBPF), Rio de Janeiro, Brazil\\
$ ^{2}$Universidade Federal do Rio de Janeiro (UFRJ), Rio de Janeiro, Brazil\\
% ...
$ ^{4}$LAPP, Universit\'{e} de Savoie, CNRS/IN2P3, Annecy-Le-Vieux, France\\
$ ^{5}$Clermont Universit\'{e}, Universit\'{e} Blaise Pascal, CNRS/IN2P3, LPC, Clermont-Ferrand, France\\
$ ^{6}$CPPM, Aix-Marseille Universit\'{e}, CNRS/IN2P3, Marseille, France\\
% ...
$ ^{33}$Universitat de Barcelona, Barcelona, Spain\\
$ ^{34}$Universidad de Santiago de Compostela, Santiago de Compostela, Spain\\
$ ^{35}$European Organization for Nuclear Research (CERN), Geneva, Switzerland\\
% ...
$ ^{38}$Nikhef National Institute for Subatomic Physics, Amsterdam, The Netherlands\\
% ...
$ ^{43}$H.H. Wills Physics Laboratory, University of Bristol, Bristol, United Kingdom\\
% ...
$ ^{48}$School of Physics and Astronomy, University of Glasgow, Glasgow, United Kingdom\\
$ ^{49}$Oliver Lodge Laboratory, University of Liverpool, Liverpool, United Kingdom\\
% ...
$ ^{n}$LIFAELS, La Salle, Universitat Ramon Llull, Barcelona, Spain\\
% ...
}
% \bigskip
% ---- LHCb Authorlist, Status 18. Dec. 2011
% ---- Number of Authors = 596
% ---- 
\end{flushleft}
%\end{document}

\cleardoublepage

% %%%%%%%%%%%%% ---------

\renewcommand{\thefootnote}{\arabic{footnote}}
\setcounter{footnote}{0}

%%%%%%%%%%%%%%%%%%%%%%%%%%%%%%%%
%%%%%  Table of Content   %%%%%%
%%%%%%%%%%%%%%%%%%%%%%%%%%%%%%%%
%%%% Uncomment next 2 lines if desired
%\tableofcontents
%\cleardoublepage

%%%%%%%%%%%%%%%%%%%%%%%%%
%%%%% Main text %%%%%%%%%
%%%%%%%%%%%%%%%%%%%%%%%%%

\pagestyle{plain} % restore page numbers for the main text
\setcounter{page}{1}
\pagenumbering{arabic}

\section{Introduction}
\label{sec:introduction}

The study of the \bbbar\ production cross-section is a powerful test of
perturbative quantum chromodynamics (pQCD) calculations. These are
available at next-to-leading order (NLO)~\cite{Nason:1987xz}
and with the fixed-order plus next-to-leading logarithms
(FONLL)~\cite{Cacciari:1998it, Cacciari:2001td} approximations.
In the NLO and FONLL calculations, the theoretical
predictions have large uncertainties arising from the choice of the renormalisation
and factorisation scales and the \textit{b}-quark mass~\cite{Cacciari:2003uh}.
Accurate measurements provide tests of the validity of the different
production models.
Recently, the LHCb collaboration measured 
the $b\bar{b}$ production cross-section in
hadron collisions using \jpsi\ from $b$ decays~\cite{Aaij:2011jh} 
and $b\to D\mu X$ decays~\cite{Aaij:2010gn}.
The two most recent measurements of the \bplus\ production cross-section in
hadron collisions have been performed by the CDF collaboration in the range
$\ptrans>6\,\gevc$ and $|y|<1$~\cite{Abulencia:2006ps},
where $\ptrans$ is the transverse momentum and
$y$ is rapidity,
and by the CMS collaboration
in the range $\ptrans>~5\,\gevc$ and $|y|<2.4$~\cite{Khachatryan:2011mk}.
This paper presents a measurement of the \bplus
production cross-section in $pp$ collisions
at a centre-of-mass energy of $\sqrt{s}=7\,\tev$ using
$34.6\pm 1.2\,\pbinv$ of data collected by the LHCb detector in 2010.
The \bplus\ mesons are reconstructed exclusively in the \bpjpsik\ mode,
with \jpsimumu.
Both the total production cross-section and the differential
cross-section, ${\rm d}\sigma/{\rm d} \ptrans$, 
as a function of the $\bplus$ transverse momentum for 
\mbox{$0<p_{\rm T}<40$\; GeV/$c$} and $2.0<y<4.5$, are measured.

The \lhcb detector~\cite{Alves:2008zz} is a single-arm forward
spectrometer covering the pseudo-rapidity range $2<\eta <5$, designed
for the study of particles containing \bquark or \cquark quarks. The
detector includes a high precision tracking system consisting of a
silicon-strip vertex detector surrounding the $pp$ interaction region,
a large-area silicon-strip detector located upstream of a dipole
magnet with a bending power of about $4{\rm\,Tm}$, and three stations
of silicon-strip detectors and straw drift-tubes placed
downstream. The combined tracking system has a momentum resolution
$\Delta p/p$ that varies from 0.4\% at 5\gevc to 0.6\% at 100\gevc,
and an impact parameter resolution of 20\mum for tracks with high
transverse momentum. Charged hadrons are identified using two
ring-imaging Cherenkov detectors. Photon, electron and hadron
candidates are identified by a calorimeter system consisting of
scintillating-pad and pre-shower detectors, an electromagnetic
calorimeter and a hadronic calorimeter. Muons are identified by a muon
system composed of alternating layers of iron and multiwire
proportional chambers. 

The LHCb detector uses a two-level trigger system, 
the first level (L0) is hardware based,
and the second level is software based high level trigger (HLT).
Here only the triggers used in this analysis are described.
At the L0 either a single muon candidate with $\ptrans$ larger than $1.4\,\gevc$
or a pair of muon candidates,
one with \ptrans\ larger than $0.56\,\gevc$ and the other with \ptrans\ larger than $0.48\,\gevc$,
is required.
Events passing these requirements are read out and sent to an event
filter farm for further selection.
In the first stage of the HLT, 
events satisfying one of the following three selections are kept:
the first one confirms the single-muon candidates from L0  
and applies a harder \ptrans\ selection at $1.8\,\gevc$; 
the second one confirms the single-muon from L0 and looks for another muon in the
event, and the third one confirms the dimuon candidates from L0.
Both the second and third selections require the dimuon invariant mass 
to be greater than $2.5\,\gevcc$.
The second stage of the HLT selects events that pass any selections of
previous stage and contain two muon candidates with an invariant mass
within 120 \mevcc\ of the known \jpsi\ mass.
To reject high-multiplicity events with a large number 
of $pp$ interactions,
a set of global event cuts (GEC) is applied on the hit multiplicities of sub-detectors.

\section{Event selection}
\label{sec:eventselection}
Candidates for $\jpsi \to \mu^+\mu^-$ decay are formed from pairs of particles with opposite charge.
Both particles are required to have a good track fit quality
($\chi^2/{\rm ndf}<4$, where ndf represents the number of degrees of
freedom in the fit), a transverse momentum $p_{\rm T}>0.7$ GeV/$c$ and to be
identified as a muon.
In addition, the muon pair is required to originate
from a common vertex ($\chi^2/{\rm ndf}<9$).
The mass of the reconstructed $\jpsi$ is required to be in the range $3.04 - 3.14\,\gevcc$.

The bachelor kaon candidates used to form \bpjpsik\ candidates 
are required to have \ptrans\ larger than 0.5 $\gevc$ and 
to have a good track fit quality ($\chi^2/{\rm ndf}<4$).
No particle identification is used in the selection of the kaon. 
A vertex fit is performed
that constrains the three daughter particles
to originate from a common point and the mass of the muon pair to
match the nominal $\jpsi$ mass.
It is required that $\chi^2/{\rm ndf}<9$ for this fit.
To further reduce the combinatorial background due to particles produced
in the primary $pp$ interaction, only candidates with a decay time larger than 0.3 ${\rm ps}$ are accepted.
Finally, the fiducial requirement \mbox{$0<p_{\rm T}<40$\; GeV/$c$}
and $2.0<y<4.5$ is applied to the \bplus candidates.

\section{Cross-section determination}
\label{sec:determination}

The differential production cross-section is measured as
\begin{equation}
\frac{{\rm d}\sigma}{{\rm d}\ptrans}=\frac{N_{\bplus}(\ptrans)}
{{\cal L}\;\epsilon_{\rm tot}(\ptrans)\;{\cal B}(\bpjpsik)\;{\cal B}(\jpsimumu)\;\Delta\ptrans},
\end{equation}
where $N_{\bplus}(\ptrans)$ is the number of reconstructed \bpjpsik\
signal events in a given \ptrans\ bin,
${\cal L}$ is the integrated luminosity,
$\epsilon_{\rm tot}(\ptrans)$ is the total efficiency,
including geometrical acceptance, reconstruction, selection and trigger
effects, ${\cal B}(\bplus\to\jpsi K^\pm)$ and ${\cal
  B}(\jpsi\to\mu^+\mu^-)$
are the branching fractions of the reconstructed decay chain~\cite{Nakamura:2010pd},
and $\Delta\ptrans$ is the \ptrans\ bin width.

\begin{figure}%[!htdp]
  \centering
  \includegraphics[width=10cm]{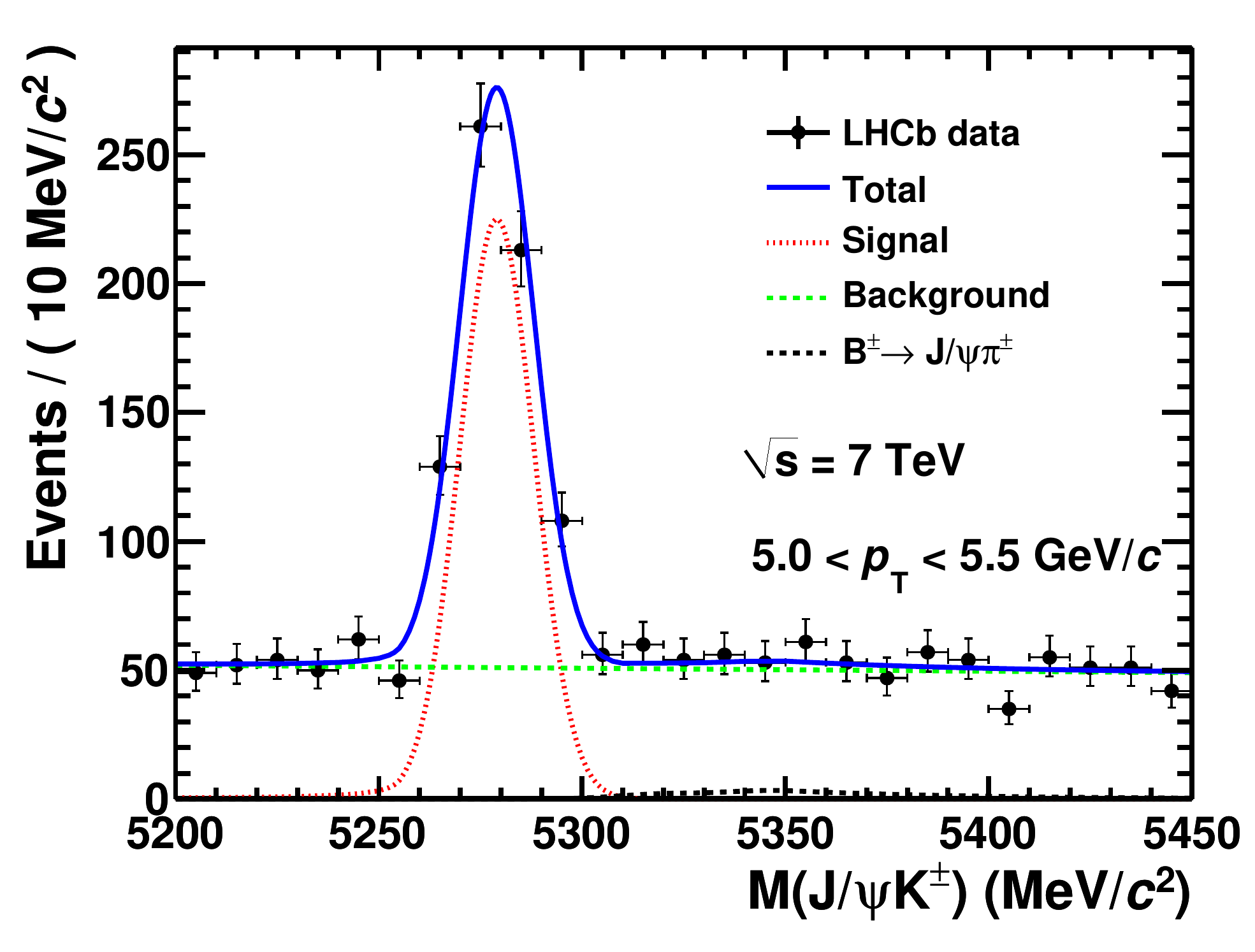}
  \caption{Invariant mass distribution of the selected \bpjpsik\
    candidates for one bin {\;\; } ($5.0<\ptrans<5.5\,\gevc$).
    The result of the fit to the model described in the text is superimposed.
  }
  \label{fig:BuMassPlot}
\end{figure}

Considering that the efficiencies depend on $p_{\rm T}$ and $y$,
we calculate the event yield in bins of these variables
using an extended unbinned maximum likelihood fit to the invariant mass
distribution of the reconstructed \bplus\ candidates in the interval
$5.15<M_{\bplus}<5.55\,\gevcc$.
We assume that the signal and background shapes only depend
on $p_{\rm T}$.
Three components are included in the fit procedure:
a \mbox{Crystal Ball function~\cite{Skwarnicki:1986ps}} to model
the signal, an exponential function to model the combinatorial
background and
a double-Crystal Ball function~\footnote{A double-Crystal Ball function has tails on both the low and high mass side of the peak with separate parameters for the two.}
to model the Cabibbo suppressed decay $\bplus\to\jpsi\pi^\pm$.
The shape of the latter component is found to
fit well the distribution of simulated events.
The ratio of the number of $\bplus\to\jpsi\pi^\pm$ candidates to that of the signal is fixed to
${\cal B}(\bpjpsipi) / {\cal B}(\bpjpsik)$ from
Ref.~\cite{Nakamura:2010pd}.
The invariant mass distribution of the selected \bpjpsik\ candidates
and the fit result for one bin ($5.0<\ptrans<5.5\,\gevc$) are shown
in Fig.~\ref{fig:BuMassPlot}. The fit returns a mass resolution of
$9.14 \pm 0.49$ \mevcc, and a mean of $5279.05 \pm 0.56$ \mevcc, where
the uncertainties are statistical only.
Summing over all $p_{\rm T}$ bins, the total
number of signal events is about 9100.

The geometrical acceptance and the reconstruction and selection
efficiencies are determined using simulated signal events.
The simulation is based on
\mbox{\textsc{Pythia}} 6.4 generator~\cite{Sjostrand:2006za}
with parameters configured for LHCb~\cite{LHCb-PROC-2011-006}.
The \mbox{\textsc{EvtGen}} package~\cite{Lange:2006za} is used to
describe the decays of the \bplus\ and \jpsi.
QED radiative corrections are modelled
using \mbox{\textsc{Photos}~\cite{Golonka:2005pn}}.
The \mbox{\textsc{Geant4}~\cite{Agostinelli:2003ge}} simulation
package is used to trace the decay products
through the detector.
Since we select events passing trigger selections that depend on $\jpsi$ properties only,
the trigger efficiency is obtained from a trigger-unbiased
data sample of \jpsi\ events that would still be triggered if the
\jpsi\ candidate were removed.
The efficiency of GEC is determined from data to be $(92.6\pm0.3)\%$,
and assumed to be independent of the \bplus\ \ptrans\ and $y$.
The total trigger efficiency is then the product
of the \jpsi\ trigger efficiency and the GEC efficiency.
The luminosity is measured using Van der Meer scans
and a beam-gas imaging method~\cite{Aaij:2011er}.
The knowledge of the absolute luminosity
scale is used to calibrate the number of tracks in the vertex detector,
which is found to be stable throughout the data-taking period
and can therefore be used to monitor the instantaneous
luminosity of the entire data sample.
The integrated luminosity of the data sample used in this analysis
is determined to be $34.6$~pb$^{-1}$.

The measurement is affected by the systematic uncertainty on the
determination of signal yields, efficiencies, branching fractions
and luminosity.

The uncertainty on the determination of the signal yields
mainly arises from the description of final state radiation
in the signal fit.
The fitted signal yield is corrected by 3.0\%,
which is estimated by comparing the fitted and generated signal yields
in the Monte Carlo simulation,
and an uncertainty of 1.5\% is assigned.
The uncertainties from the effects of
the Cabibbo-suppressed background,
multiple candidates and mass fit range are found to be negligible.

The uncertainties on the efficiencies arise from trigger ($0.5-6.0$\% depending on the bin),
tracking ($3.9-4.4$\% depending on the bin), muon identification (2.5\%)~\cite{Aaij:2011jh}
and the vertex fit quality cut (1.0\%).
The trigger systematic uncertainty
has been evaluated by measuring the trigger efficiency in the
simulation using a trigger-unbiased data sample of simulated \jpsi\ events.
The tracking uncertainty includes two components: the first one
is the differences in track reconstruction
efficiency between data and simulation,
estimated with a tag and probe method~\cite{Aaij:2010nx}
using $\jpsi\rightarrow\mu^+\mu^-$ events;
the second is due to the 2\% uncertainty on the hadronic interaction length
of the detector used in the simulation.
The uncertainties from the effects of GEC,
$\jpsi$ mass window cut
and inter-bin cross-feed
are found to be negligible.
The uncertainty due to the choice of \ptrans\ binning is estimated to
be smaller than 2.0\%.

The product of ${\cal B}(\bplus\to\jpsi K^\pm)$ and ${\cal
  B}(\jpsi\to\mu^+\mu^-)$ is calculated to be $(6.01 \pm 0.20)\times
10^{-5}$, by taking their values from Ref.~\cite{Nakamura:2010pd}
with their correlations taken into account.

The absolute luminosity scale is measured
with a $3.5\%$ uncertainty~\cite{Aaij:2011er}, dominated by the beam current uncertainty.

\section{Results and conclusion}
\label{sec:results}
\begin{figure}%[!bhtdp]
\centering
\includegraphics[width=7.5cm]{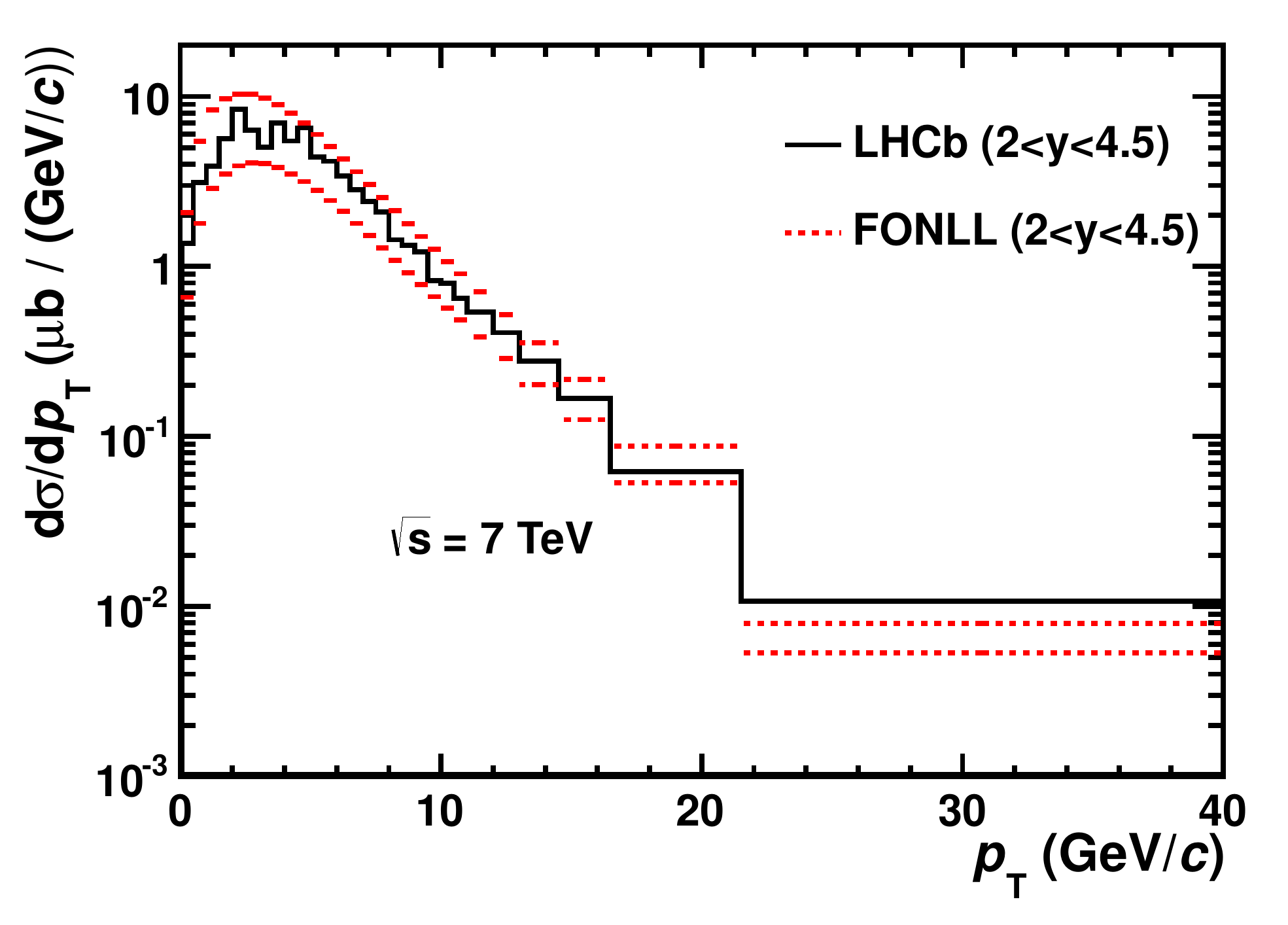}
\includegraphics[width=7.5cm]{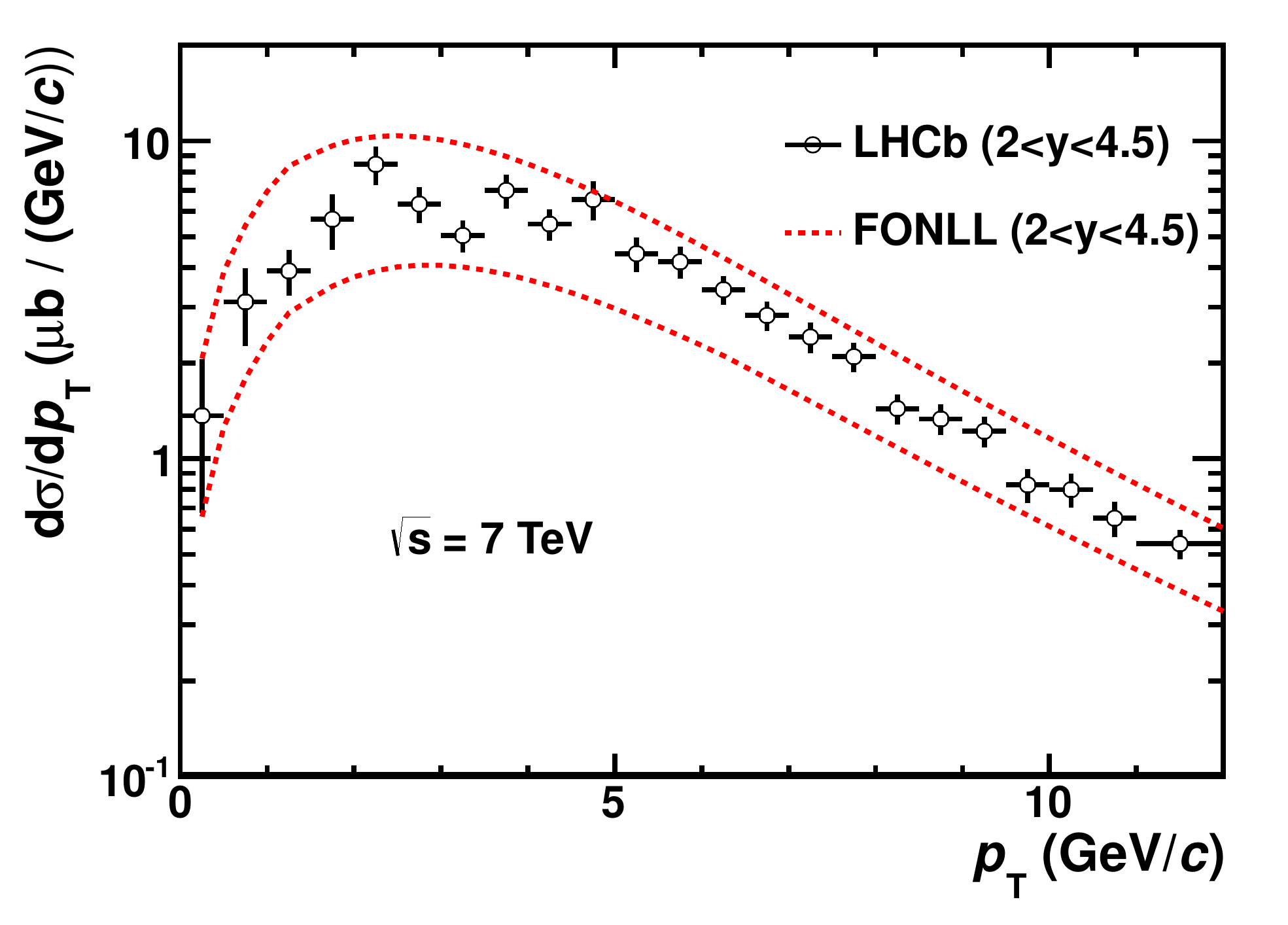}
\caption{Differential production cross-section as a function of the
  \bplus\ transverse momentum.
  The left plot shows the full $p_{\rm T}$ range,
  the right plot shows a zoom of the $p_{\rm T}$ range of $0-12$ \gevc.
  The histogram (left) and the open circles with error bars (right) are
  the measurements. The red dashed lines in both plots
  are the upper and lower uncertainty limits of the \textsc{FONLL}
  computation.
  A hadronisation fraction $f_{\bar{b}\to B^+}$ of
  $(40.1\pm1.3)$\%~\cite{Nakamura:2010pd} is
  assumed to fix the overall scale.
  The uncertainty of the \textsc{FONLL} computation includes the
  uncertainties of the $b$-quark mass,
  renormalisation and factorisation scales, and CTEQ~6.6 PDF.}
\label{fig:fonll}
\end{figure}

The measured \bplus\ differential production cross-section
in bins of \ptrans\ for $2.0<y<4.5$
is given in Table~\ref{tab:xsecresults}.
This result is compared with
a \textsc{FONLL} prediction~\cite{Cacciari:1998it, Cacciari:2001td} in Fig.~\ref{fig:fonll}.
A hadronisation fraction $f_{\bar{b}\to B^+}$ of $(40.1\pm1.3)$\%~\cite{Nakamura:2010pd}
is assumed to fix the overall
scale of \textsc{FONLL}.
The uncertainty of the \textsc{FONLL} computation includes the uncertainties on the $b$-quark mass,
renormalisation and factorisation scales, and CTEQ~6.6~\cite{Nadolsky:2008zw}
Parton Density Functions (PDF).
Good agreement is observed between data and the FONLL prediction.
The integrated cross-section is
\begin{equation}
  \sigma(pp \to \bplus X,\;\mbox{$0<p_{\rm T}<40$\; GeV/$c$},\; 2.0<y<4.5) = \xsresult.\nonumber
\label{eq:result}
\end{equation}
This is the first measurement of $\bplus$ production in the forward region at
$\sqrt{s}$ = 7 TeV.

\renewcommand{\arraystretch}{1.2}
\begin{table}[!htdp]
\tabcolsep 4mm
\begin{center}
  \caption{\label{tab:xsecresults} Differential \bplus\
    production cross-section in
    bins of $\ptrans$ for $2.0<y<4.5$. The first and second quoted
    uncertainties are statistical and
    systematic, respectively.}
  \begin{tabular}{@{}cc|cc@{}}
\toprule
$\ptrans$ $(\gevc)$ & ${\rm d \sigma}/{{\rm d}\ptrans}$ $({\mub}/(\gevc))$ & $\ptrans$ $(\gevc)$ & ${\rm d \sigma}/{{\rm d}\ptrans}$ $({\mub}/(\gevc))$\\
\midrule
$ 0.0 - 0.5 $ & 1.37 $\pm$ 0.68 $\pm$ 0.13 & $ 7.0 - 7.5 $ & 2.42 $\pm$ 0.20 $\pm$ 0.18 \\
$ 0.5 - 1.0 $ & 3.12 $\pm$ 0.82 $\pm$ 0.24 & $ 7.5 - 8.0 $ & 2.09 $\pm$ 0.16 $\pm$ 0.15 \\
$ 1.0 - 1.5 $ & 3.90 $\pm$ 0.57 $\pm$ 0.29 & $ 8.0 - 8.5 $ & 1.44 $\pm$ 0.11 $\pm$ 0.11 \\
$ 1.5 - 2.0 $ & 5.67 $\pm$ 1.05 $\pm$ 0.43 & $ 8.5 - 9.0 $ & 1.33 $\pm$ 0.11 $\pm$ 0.10 \\
$ 2.0 - 2.5 $ & 8.44 $\pm$ 1.00 $\pm$ 0.64 & $ 9.0 - 9.5 $ & 1.22 $\pm$ 0.10 $\pm$ 0.09 \\
$ 2.5 - 3.0 $ & 6.33 $\pm$ 0.66 $\pm$ 0.48 & $ 9.5 - 10.0 $ & 0.83 $\pm$ 0.08 $\pm$ 0.06 \\
$ 3.0 - 3.5 $ & 5.04 $\pm$ 0.45 $\pm$ 0.38 & $ 10.0 - 10.5 $ & 0.80 $\pm$ 0.08 $\pm$ 0.06 \\
$ 3.5 - 4.0 $ & 6.99 $\pm$ 0.68 $\pm$ 0.52 & $ 10.5 - 11.0 $ & 0.65 $\pm$ 0.07 $\pm$ 0.05 \\
$ 4.0 - 4.5 $ & 5.48 $\pm$ 0.47 $\pm$ 0.41 & $ 11.0 - 12.0 $ & 0.54 $\pm$ 0.04 $\pm$ 0.04 \\
$ 4.5 - 5.0 $ & 6.54 $\pm$ 0.79 $\pm$ 0.49 & $ 12.0 - 13.0 $ & 0.41 $\pm$ 0.04 $\pm$ 0.03 \\
$ 5.0 - 5.5 $ & 4.42 $\pm$ 0.44 $\pm$ 0.33 & $ 13.0 - 14.5 $ & 0.28 $\pm$ 0.02 $\pm$ 0.02 \\
$ 5.5 - 6.0 $ & 4.16 $\pm$ 0.37 $\pm$ 0.31 & $ 14.5 - 16.5 $ & 0.17 $\pm$ 0.02 $\pm$ 0.01 \\
$ 6.0 - 6.5 $ & 3.40 $\pm$ 0.24 $\pm$ 0.25 & $ 16.5 - 21.5 $ & 0.062 $\pm$ 0.005 $\pm$ 0.005 \\
$ 6.5 - 7.0 $ & 2.82 $\pm$ 0.22 $\pm$ 0.21 & $ 21.5 - 40.0 $ & 0.011 $\pm$ 0.001 $\pm$ 0.001 \\
\bottomrule
\end{tabular}
\end{center}
\end{table}
\renewcommand{\arraystretch}{1.}

% No need to include this in an analysis note!
\section*{Acknowledgements}

\noindent We express our gratitude to our colleagues in the CERN accelerator
departments for the excellent performance of the LHC. We thank the
technical and administrative staff at CERN and at the LHCb institutes,
and acknowledge support from the National Agencies: CAPES, CNPq,
FAPERJ and FINEP (Brazil); CERN; NSFC (China); CNRS/IN2P3 (France);
BMBF, DFG, HGF and MPG (Germany); SFI (Ireland); INFN (Italy); FOM and
NWO (The Netherlands); SCSR (Poland); ANCS (Romania); MinES of Russia and
Rosatom (Russia); MICINN, XuntaGal and GENCAT (Spain); SNSF and SER
(Switzerland); NAS Ukraine (Ukraine); STFC (United Kingdom); NSF
(USA). We also acknowledge the support received from the ERC under FP7
and the Region Auvergne.

\clearpage
\bibliographystyle{LHCb}
\bibliography{lhcb-paper-2011-043}

\ifx\mcitethebibliography\mciteundefinedmacro
\PackageError{LHCb.bst}{mciteplus.sty has not been loaded}
{This bibstyle requires the use of the mciteplus package.}\fi
\providecommand{\href}[2]{#2}
\begin{mcitethebibliography}{10}
\mciteSetBstSublistMode{n}
\mciteSetBstMaxWidthForm{subitem}{\alph{mcitesubitemcount})}
\mciteSetBstSublistLabelBeginEnd{\mcitemaxwidthsubitemform\space}
{\relax}{\relax}

\bibitem{Nason:1987xz}
P.~Nason, S.~Dawson, and R.~K. Ellis, \ifthenelse{\boolean{articletitles}}{{\it
  {The total cross-section for the production of heavy quarks in hadronic
  collisions}}, }{}\href{http://dx.doi.org/10.1016/0550-3213(88)90422-1}{Nucl.
  Phys. {\bf B303} (1988) 607}\relax
\mciteBstWouldAddEndPuncttrue
\mciteSetBstMidEndSepPunct{\mcitedefaultmidpunct}
{\mcitedefaultendpunct}{\mcitedefaultseppunct}\relax
\EndOfBibitem
\bibitem{Cacciari:1998it}
M.~Cacciari, M.~Greco, and P.~Nason, \ifthenelse{\boolean{articletitles}}{{\it
  {The $p_T$ spectrum in heavy flavor hadroproduction}},
  }{}\href{http://dx.doi.org/10.1088/1126-6708/1998/05/007}{JHEP {\bf 05}
  (1998) 007}, \href{http://xxx.lanl.gov/abs/hep-ph/9803400}{{\tt
  arXiv:hep-ph/9803400}}\relax
\mciteBstWouldAddEndPuncttrue
\mciteSetBstMidEndSepPunct{\mcitedefaultmidpunct}
{\mcitedefaultendpunct}{\mcitedefaultseppunct}\relax
\EndOfBibitem
\bibitem{Cacciari:2001td}
M.~Cacciari, S.~Frixione, and P.~Nason,
  \ifthenelse{\boolean{articletitles}}{{\it {The $p_T$ spectrum in heavy flavor
  photoproduction}},
  }{}\href{http://dx.doi.org/10.1088/1126-6708/2001/03/006}{JHEP {\bf 03}
  (2001) 006}, \href{http://xxx.lanl.gov/abs/hep-ph/0102134}{{\tt
  arXiv:hep-ph/0102134}}\relax
\mciteBstWouldAddEndPuncttrue
\mciteSetBstMidEndSepPunct{\mcitedefaultmidpunct}
{\mcitedefaultendpunct}{\mcitedefaultseppunct}\relax
\EndOfBibitem
\bibitem{Cacciari:2003uh}
M.~Cacciari {\em et~al.}, \ifthenelse{\boolean{articletitles}}{{\it {QCD
  analysis of first $b$ cross-section data at 1.96 TeV}},
  }{}\href{http://dx.doi.org/10.1088/1126-6708/2004/07/033}{JHEP {\bf 07}
  (2004) 033}, \href{http://xxx.lanl.gov/abs/hep-ph/0312132}{{\tt
  arXiv:hep-ph/0312132}}\relax
\mciteBstWouldAddEndPuncttrue
\mciteSetBstMidEndSepPunct{\mcitedefaultmidpunct}
{\mcitedefaultendpunct}{\mcitedefaultseppunct}\relax
\EndOfBibitem
\bibitem{Aaij:2011jh}
LHCb collaboration, R.~Aaij {\em et~al.},
  \ifthenelse{\boolean{articletitles}}{{\it {Measurement of $J/\psi$ production
  in pp collisions at $\sqrt{s}$=7 TeV}},
  }{}\href{http://dx.doi.org/10.1140/epjc/s10052-011-1645-y}{Eur. Phys. J. {\bf
  C71} (2011) 1645}, \href{http://xxx.lanl.gov/abs/1103.0423}{{\tt
  arXiv:1103.0423}}\relax
\mciteBstWouldAddEndPuncttrue
\mciteSetBstMidEndSepPunct{\mcitedefaultmidpunct}
{\mcitedefaultendpunct}{\mcitedefaultseppunct}\relax
\EndOfBibitem
\bibitem{Aaij:2010gn}
LHCb collaboration, R.~Aaij {\em et~al.},
  \ifthenelse{\boolean{articletitles}}{{\it {Measurement of $\sigma(pp \to b
  \bar{b} X)$ at $\sqrt{s}=7$ TeV in the forward region}},
  }{}\href{http://dx.doi.org/10.1016/j.physletb.2010.10.010}{Phys. Lett. {\bf
  B694} (2010) 209}, \href{http://xxx.lanl.gov/abs/1009.2731}{{\tt
  arXiv:1009.2731}}\relax
\mciteBstWouldAddEndPuncttrue
\mciteSetBstMidEndSepPunct{\mcitedefaultmidpunct}
{\mcitedefaultendpunct}{\mcitedefaultseppunct}\relax
\EndOfBibitem
\bibitem{Abulencia:2006ps}
CDF collaboration, A.~Abulencia {\em et~al.},
  \ifthenelse{\boolean{articletitles}}{{\it {Measurement of the $B^+$
  production cross-section in $p\bar{p}$ collisions at $\sqrt{s} = 1960$~GeV}},
  }{}\href{http://dx.doi.org/10.1103/PhysRevD.75.012010}{Phys. Rev. {\bf D75}
  (2007) 012010}, \href{http://xxx.lanl.gov/abs/hep-ex/0612015}{{\tt
  arXiv:hep-ex/0612015}}\relax
\mciteBstWouldAddEndPuncttrue
\mciteSetBstMidEndSepPunct{\mcitedefaultmidpunct}
{\mcitedefaultendpunct}{\mcitedefaultseppunct}\relax
\EndOfBibitem
\bibitem{Khachatryan:2011mk}
CMS collaboration, V.~Khachatryan {\em et~al.},
  \ifthenelse{\boolean{articletitles}}{{\it {Measurement of the $B^+$
  production cross section in $pp$ Collisions at $\sqrt{s} = 7$~TeV}},
  }{}\href{http://dx.doi.org/10.1103/PhysRevLett.106.112001}{Phys. Rev. Lett.
  {\bf 106} (2011) 112001}, \href{http://xxx.lanl.gov/abs/1101.0131}{{\tt
  arXiv:1101.0131}}\relax
\mciteBstWouldAddEndPuncttrue
\mciteSetBstMidEndSepPunct{\mcitedefaultmidpunct}
{\mcitedefaultendpunct}{\mcitedefaultseppunct}\relax
\EndOfBibitem
\bibitem{Alves:2008zz}
LHCb collaboration, A.~A. Alves~Jr. {\em et~al.},
  \ifthenelse{\boolean{articletitles}}{{\it {The \lhcb detector at the LHC}},
  }{}\href{http://dx.doi.org/10.1088/1748-0221/3/08/S08005}{JINST {\bf 3}
  (2008) S08005}\relax
\mciteBstWouldAddEndPuncttrue
\mciteSetBstMidEndSepPunct{\mcitedefaultmidpunct}
{\mcitedefaultendpunct}{\mcitedefaultseppunct}\relax
\EndOfBibitem
\bibitem{Nakamura:2010pd}
Particle Data Group, K.~Nakamura {\em et~al.},
  \ifthenelse{\boolean{articletitles}}{{\it {Review of particle physics}},
  }{}\href{http://dx.doi.org/10.1088/0954-3899/37/7A/075021/}{J. Phys. {\bf
  G37} (2010) 075021}\relax
\mciteBstWouldAddEndPuncttrue
\mciteSetBstMidEndSepPunct{\mcitedefaultmidpunct}
{\mcitedefaultendpunct}{\mcitedefaultseppunct}\relax
\EndOfBibitem
\bibitem{Skwarnicki:1986ps}
T.~Skwarnicki, \ifthenelse{\boolean{articletitles}}{{\it {A study of the
  radiative cascade transitions between the $\Upsilon^\prime$ and $\Upsilon$
  resonances}}, }{} Ph.D. Thesis, DESY-F31-86-02 (1986)\relax
\mciteBstWouldAddEndPuncttrue
\mciteSetBstMidEndSepPunct{\mcitedefaultmidpunct}
{\mcitedefaultendpunct}{\mcitedefaultseppunct}\relax
\EndOfBibitem
\bibitem{Sjostrand:2006za}
T.~Sj\"{o}strand, S.~Mrenna, and P.~Skands,
  \ifthenelse{\boolean{articletitles}}{{\it {PYTHIA 6.4 physics and manual}},
  }{}\href{http://dx.doi.org/10.1088/1126-6708/2006/05/026}{JHEP {\bf 05}
  (2006) 026}, \href{http://xxx.lanl.gov/abs/hep-ph/0603175}{{\tt
  arXiv:hep-ph/0603175}}\relax
\mciteBstWouldAddEndPuncttrue
\mciteSetBstMidEndSepPunct{\mcitedefaultmidpunct}
{\mcitedefaultendpunct}{\mcitedefaultseppunct}\relax
\EndOfBibitem
\bibitem{LHCb-PROC-2011-006}
M.~Clemencic {\em et~al.}, \ifthenelse{\boolean{articletitles}}{{\it {The \lhcb
  Simulation Application, Gauss: Design, Evolution and Experience}},
  }{}\href{http://dx.doi.org/10.1088/1742-6596/331/3/032023}{{Journal of
  Physics: Conference Series} {\bf 331} (2011), no.~3 032023}\relax
\mciteBstWouldAddEndPuncttrue
\mciteSetBstMidEndSepPunct{\mcitedefaultmidpunct}
{\mcitedefaultendpunct}{\mcitedefaultseppunct}\relax
\EndOfBibitem
\bibitem{Lange:2006za}
D.~J. Lange, \ifthenelse{\boolean{articletitles}}{{\it {The EVTGEN particle
  decay simulation package}},
  }{}\href{http://dx.doi.org/10.1016/S0168-9002(01)00089-4}{Nucl. Instrum.
  Methods {\bf A462} (2001) 152}\relax
\mciteBstWouldAddEndPuncttrue
\mciteSetBstMidEndSepPunct{\mcitedefaultmidpunct}
{\mcitedefaultendpunct}{\mcitedefaultseppunct}\relax
\EndOfBibitem
\bibitem{Golonka:2005pn}
P.~Golonka and Z.~Was, \ifthenelse{\boolean{articletitles}}{{\it {PHOTOS Monte
  Carlo: A precision tool for QED corrections in $Z$ and $W$ decays}},
  }{}\href{http://dx.doi.org/10.1140/epjc/s2005-02396-4}{Eur. Phys. J. {\bf
  C45} (2006) 97}, \href{http://xxx.lanl.gov/abs/hep-ph/0506026}{{\tt
  arXiv:hep-ph/0506026}}\relax
\mciteBstWouldAddEndPuncttrue
\mciteSetBstMidEndSepPunct{\mcitedefaultmidpunct}
{\mcitedefaultendpunct}{\mcitedefaultseppunct}\relax
\EndOfBibitem
\bibitem{Agostinelli:2003ge}
\textsc{Geant4} collaboration, S.~Agostinelli {\em et~al.},
  \ifthenelse{\boolean{articletitles}}{{\it {GEANT4: a simulation toolkit}},
  }{}\href{http://dx.doi.org/10.1016/S0168-9002(03)01368-8}{Nucl. Instrum.
  Methods {\bf A506} (2003) 250}\relax
\mciteBstWouldAddEndPuncttrue
\mciteSetBstMidEndSepPunct{\mcitedefaultmidpunct}
{\mcitedefaultendpunct}{\mcitedefaultseppunct}\relax
\EndOfBibitem
\bibitem{Aaij:2011er}
LHCb collaboration, R.~Aaij {\em et~al.},
  \ifthenelse{\boolean{articletitles}}{{\it {Absolute luminosity measurements
  with the LHCb detector at the LHC}},
  }{}\href{http://dx.doi.org/10.1088/1748-0221/7/01/P01010}{JINST {\bf 7}
  (2012) P01010}, \href{http://xxx.lanl.gov/abs/1110.2866}{{\tt
  arXiv:1110.2866}}\relax
\mciteBstWouldAddEndPuncttrue
\mciteSetBstMidEndSepPunct{\mcitedefaultmidpunct}
{\mcitedefaultendpunct}{\mcitedefaultseppunct}\relax
\EndOfBibitem
\bibitem{Aaij:2010nx}
LHCb collaboration, R.~Aaij {\em et~al.},
  \ifthenelse{\boolean{articletitles}}{{\it {Prompt $K_s^0$ production in $pp$
  collisions at $\sqrt{s}=0.9$ TeV}},
  }{}\href{http://dx.doi.org/10.1016/j.physletb.2010.08.055}{Phys. Lett. {\bf
  B693} (2010) 69}, \href{http://xxx.lanl.gov/abs/1008.3105}{{\tt
  arXiv:1008.3105}}\relax
\mciteBstWouldAddEndPuncttrue
\mciteSetBstMidEndSepPunct{\mcitedefaultmidpunct}
{\mcitedefaultendpunct}{\mcitedefaultseppunct}\relax
\EndOfBibitem
\bibitem{Nadolsky:2008zw}
P.~M. Nadolsky {\em et~al.}, \ifthenelse{\boolean{articletitles}}{{\it
  {Implications of CTEQ global analysis for collider observables}},
  }{}\href{http://dx.doi.org/10.1103/PhysRevD.78.013004}{Phys. Rev. {\bf D78}
  (2008) 013004}, \href{http://xxx.lanl.gov/abs/0802.0007}{{\tt
  arXiv:0802.0007}}\relax
\mciteBstWouldAddEndPuncttrue
\mciteSetBstMidEndSepPunct{\mcitedefaultmidpunct}
{\mcitedefaultendpunct}{\mcitedefaultseppunct}\relax
\EndOfBibitem
\end{mcitethebibliography}

\end{document}